\documentclass[sigconf]{acmart}

\AtBeginDocument{%
  }
\DeclareMathAlphabet{\mathmybb}{U}{bbold}{m}{n}
\newcommand{\1}{\mathmybb{1}}
\usepackage{amsmath}
\usepackage{colortbl}
\usepackage{subcaption} 
\usepackage{graphicx}
\definecolor{Gray}{gray}{0.9}
\usepackage{soul}

\begin{document}
 \fancyhead{}
\title{DCN$^2$: Interplay of \emph{Implicit} Collision Weights and \\\emph{Explicit} Cross Layers for Large-Scale Recommendation}


\author{Bla\v{z} \v{S}krlj}
\affiliation{%
  \institution{Teads}
  \country{}}
\email{blaz.skrlj@teads.com}

\author{Yonatan Karni}
\affiliation{%
  \institution{Teads}
  \country{}}
\email{yonatan.karni@teads.com}

\author{Grega Ga\v{s}per\v{s}i\v{c}}
\affiliation{%
  \institution{Teads}
  \country{}}
\email{grega.gaspersic @teads.com}

\author{Bla\v{z} Mramor}
\affiliation{%
  \institution{Teads}
  \country{}}
\email{blaz.mramor@teads.com}

\author{Yulia Stolin}
\affiliation{%
  \institution{Teads}
  \country{}}
\email{yulia.stolin@teads.com}

\author{Martin Jakomin}
\affiliation{%
  \institution{Teads}
  \country{}}
\email{martin.jakomin@teads.com}

\author{Jasna Urban\v{c}i\v{c}}
\affiliation{%
  \institution{Teads}
  \country{}}
\email{jasna.urbancic@teads.com}

\author{Yuval Dishi}
\affiliation{%
  \institution{Teads}
  \country{}}
\email{yuval.dishi@teads.com}

\author{Natalia Silberstein}
\affiliation{%
  \institution{Teads}
  \country{}}
\email{natalia.silberstein@teads.com}

\author{Ophir Friedler}
\affiliation{%
  \institution{Teads}
  \country{}}
\email{ophir.friedler@teads.com}

\author{Assaf Klein}
\affiliation{%
  \institution{Teads}
  \country{}}
\email{assaf.klein@teads.com}

\renewcommand{\shortauthors}{\v{S}krlj et al.}

\begin{abstract}
The Deep and Cross architecture (DCNv2) is a robust production baseline and is integral to numerous real-life recommender systems. Its inherent efficiency and ability to model interactions often result in models that are both simpler and highly competitive compared to more computationally demanding alternatives, such as Deep FFMs. In this work, we introduce three significant algorithmic improvements to the DCNv2 architecture, detailing their formulation and behavior at scale. The enhanced architecture we refer to as DCN$^2$ is actively used in a live recommender system, processing over 0.5 billion predictions per second across diverse use cases where it out-performed DCNv2, both offline and online (a/b tests). These improvements effectively address key limitations observed in the DCNv2, including information loss in Cross layers, implicit management of collisions through learnable lookup-level weights, and explicit modeling of pairwise similarities with a custom layer that emulates FFMs' behavior. The superior performance of DCN$^2$ is also demonstrated on four publicly available benchmark data sets.
\end{abstract}




\maketitle

\section{Introduction}
\label{sec:intro}
\begin{figure}[b!]
    \centering
    \includegraphics[width=1.0\linewidth]{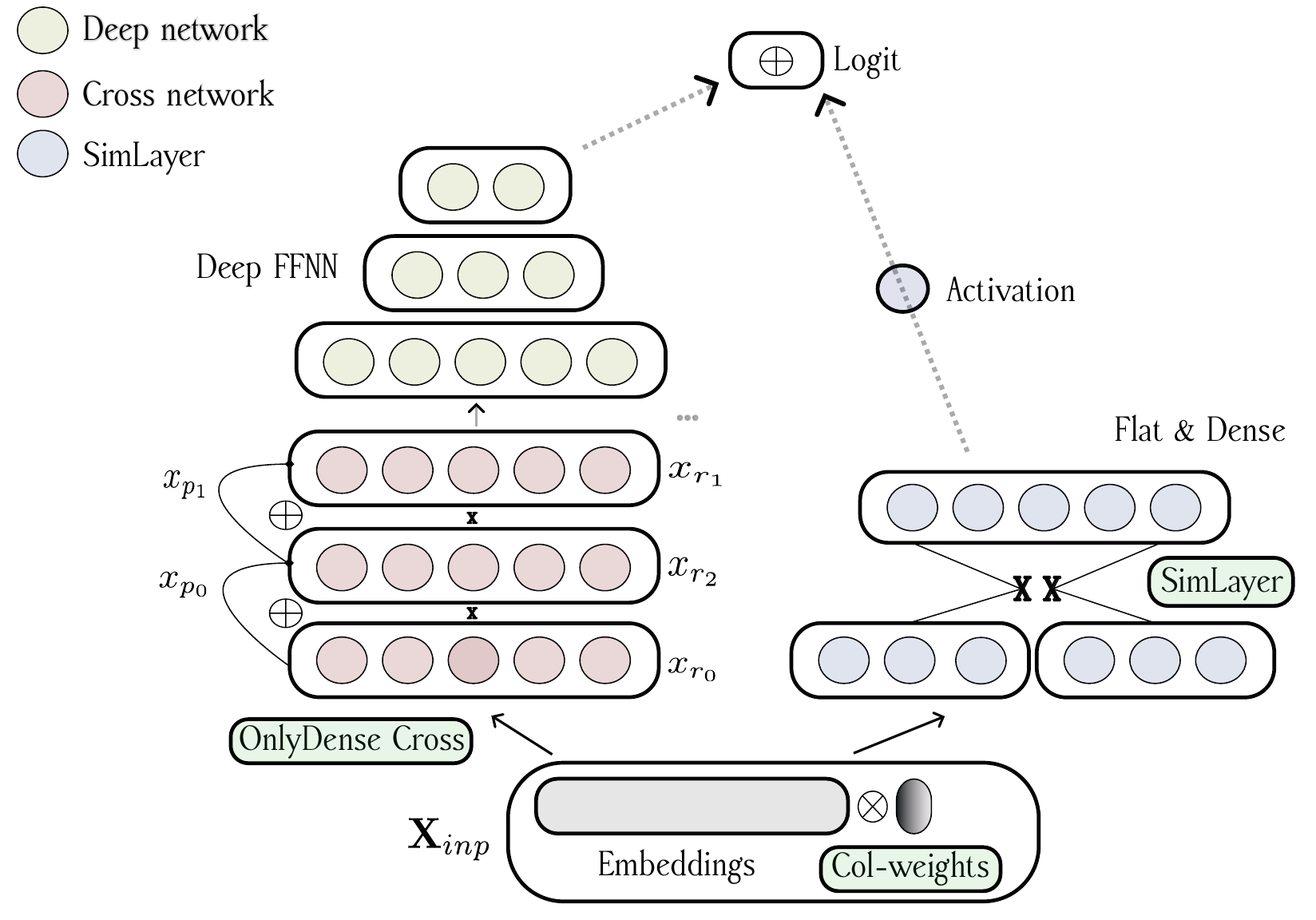}
    \caption{Architecture of DCN$^2$. Parts marked in green denote additions on top of DCNv2~\cite{Wang2021} presented in this paper.}
    \label{fig:overviewimg}
\end{figure}
Contemporary \emph{recommender systems} are composed of various components, with \textbf{factorization-machine-based models} playing a critical role in large-scale applications. These machine learning models are favored for their ability to deliver fast inference while maintaining strong predictive performance. While the original \emph{factorization machines} (FMs)~\cite{rendle2010factorization} served as a generalization of logistic regression, advancements have led to more sophisticated architectures, such as Deep and Cross Networks (DCN)~\cite{wang2017deep}. The second iteration, DCNv2~\cite{Wang2021}, successfully addressed scalability and performance issues, maintaining its status as a leading neural network architecture for tasks like click-through rate and conversion prediction.
Despite DCNv2's capabilities, certain algorithmic aspects remain sub-optimal. It lacks a specific mechanism for \emph{collision} management, i.e. accounting for effects of multiple items having same representation~\cite{freksen2018fully}, often deferring this responsibility to external hashing logic. Moreover, although \textbf{Cross layers} are theoretically advantageous for modeling higher-order interactions, they may suffer from information loss due to the compression inherent to each intermediary projection. Additionally, the algorithmic bias in Cross networks differs from that of Field-aware Factorization Machines (FFMs), where performance trade-offs have been extensively optimized (explicit interactions). Contributions of this paper are multifold:
    We present DCN$^2$, an architecture building on DCNv2, designed to address issues related to collision handling, algorithmic bias in pairwise interactions, and lossiness of Cross layers; the \textbf{collision handling mechanism} ensures lookup-level weights can modulate impact of new collisions, pairwise interactions are explicitly handled as we show that DCNv2 does not entail all, and the cross layer itself was improved to capture more fine-grained interaction patterns and lose less information layer-to-layer.
    The performance of DCN$^2$ is evaluated against DCNv2 using four real-life benchmark datasets, demonstrating competitive results, as well as within a live recommender that handles more than 0.5 billion pps, where the new architecture significantly out-performs the existing DCNv2 based model (+3\% RPM).
    We outline the deployment pathway for DCN$^2$, which now serves as the flagship model, processing over 0.5 billion predictions per second across diverse applications.
    Insights and battle scars from optimizing online performance in a live recommender are discussed, including inference optimization, model migration, and handling of multi-value features.

\section{Related Work}
\label{sec:related}
Web-scale recommender systems have in recent years converged to a combination of predictive models, combined with filtering/sorting tailored to a given use case~\cite{fan2024recommender}. \textbf{Factorization machine} based approaches became the gold standard methodology for estimating click-through rates, conversions or similar. This branch of methods extends the commonly used logistic regression based classifiers by incorporating either explicit or implicit notion of feature interaction. One significant extension to the paradigm of myopic LR are factorization machines (FM)~\cite{rendle2010factorization}. This branch of approaches attempts to explicitly or implicitly model interactions between features (or directly their values). Recent extensions of this paradigm include higher-order factorization machines~\cite{blondel2016higher}, field-aware factorization machines~\cite{juan2016field} and Deep FFMs~\cite{vskrlj2024bag}. Further, with the advent of GPU-based compute, deep learning-based approaches started to dominate in this domain. Approaches such as AutoInt~\cite{song2019autoint} exploit the attention layer-like mechanism to automatically unveil relevant interactions. Recent approaches also investigated use of Graph Neural Network-like patterns for building factorization machines~\cite{wu2025graphfm}. Computational efficiency, however, remains one of the key aspects of factorization machine based approaches in real-life production. One of the methods that moved the boundary between both compute efficiency and predictive performance are the Deep\&Cross networks (DCNv2~\cite{Wang2021}). This type of neural network architecture introducess the \emph{Cross} layers, a mechanism that models higher order interactions via iterative building of the latent space between the initial embeddings (inputs). The DCNv2 architecture remains one of the top-performing, off-the-shelf solutions for building larger-scale recommender systems. Insights obtained as part of research related to DCNv2 have been integrated into widely used deep learning frameworks such as Tensorflow~\cite{tensorflow2015-whitepaper}.
Simultaneously to research on neural network architectures, inspection of the impact of \emph{collisions}, i.e.  same-representation multi-item occurrences, became a lively research area of its own. Approaches that mitigate negative impacts of collisions are discussed next. One prominent direction is by incorporating \emph{semantic IDs}, high-level identifiers that preserve semantic meaning of items~\cite{zheng2025enhancing}. Alternatively, approaches that parametrize the hash space to mitigate collision impact (i.e. learning-to-collide) have also shown promising results~\cite{ghaemmaghami2022learning}. Real-life systems such as Monolith~\cite{liu2022monolith} implement the collisionless principle -- re-hashing of items if required to preserve collision-free embedding space. Mitigating the \textbf{issue of collisions} is paramount when designing a modern recommender system, as otherwise loss of predictive performance is expected regardless of architecture details~\cite{li2024embedding}. This work sources from both areas of research and proposes an architecture that operates at the level of DCNv2, yet is better suited for high-velocity real-life data streams where negative impact of collisions becomes prominent in time.

\section{DCN$^2$ architecture}
\label{sec:architecture}
This section focuses on DCN$^2$ architecture, key components, relation to DCNv2 and its computational complexity. Schematic overview of the architecture with main highlighted differences to DCNv2 is shown in Figure~\ref{fig:overviewimg}.

\subsection{Collision-weighted lookups}
The first algorithmic improvement stems from the observation that, regardless of the lookup table size, collision rates are high for realistic data streams. Instead of resorting to dedicated hashing schemes as many existing approaches, we propose an alternative that is more intrinsic to the neural network itself. Instead of explicitly attempting to reduce collisions, we introduce a weight for each lookup. The weight is initially set to 1.0, implying it has no impact on the existing lookup scheme. However, in time, weights per lookups change with regards to either new items, or collision-induced effects. By being able to quickly modify importance of each lookup, we noticed the neural network is able to adapt to new parts of the stream faster, yielding significant performance improvement in terms of relative information gain (RIG)~\cite{yi2013predictive}. For example, changes of publishers in the data stream (appearing/disappearing) require new embeddings, implying they will be shared at representation level eventually. The collision weights are formulated as follows\footnote{This idea was inspired by notion of polysemanticity in LLMs~\cite{scherlis2022polysemanticity}.}. Let $\textbf{X}_e \in \mathbb{R}^{b \times d}$ represent the commonly considered item embedding table; $b$ is the set of possible lookups (table's capacity) and $d$ the embedding dimension. Next, consider a table with one additional dimension ($\mathbf{X} \in \mathbb{R}^{b \times (d+1)}$) that has the following initialization structure:
\begin{equation*}
\mathbf{X}_{ec} =
\begin{cases}
    \mathbf{X}[:, 1:d] = \mathbf{X}[:, 1:d] ; -\omega \leq \mathcal{N}(\mu, \sigma^2) \leq \omega,\\
    \mathbf{X}[:, d+1] = \mathmybb{1}
\end{cases}
\end{equation*}
where $\mathbf{X}_{ec}$ represents the new embedding matrix, and $\mathcal{N}$ the standard normal distribution. $\omega$ denotes allowed bound for the values. Having defined the partitioned matrix, we further define the main lookup operation as follows.
\begin{equation*}
    \mathbf{X}_{\textrm{inp}} = \mathbf{X}_{ec}[:,\dots d] \odot \mathbf{X}_{ec}[:,d+1],
\end{equation*}
where $\odot$ denotes row-scalar products between the embedding matrix and the weight vector ($\1$). By definition, the initial structure can be interpreted as a generalization of the embedding matrix where all collision weights are equal to one, thus no impact of the collision weights is expected. However, these weights are differentiable, and are subject to updates alongside the main embedding space (they change in time, reducing the importance of frequently colliding items). The additional computational overhead is bounded by $\mathcal{O}(|X|)$ additional multiplications per lookup (scalar-matrix multiplication), and is in practice not significant (relative to the remainder of the architecture, these multiplications represent the minority of compute time).

\subsection{\emph{onlydense} layer as an alternative to Cross}
\begin{figure}[t!]
    \centering
    \includegraphics[width=.9\linewidth]{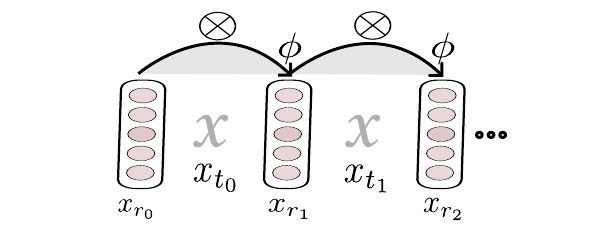}
    \vspace{-0.5cm}
    \caption{Example with two hidden \emph{onlydense} layers.} \vspace{-0.5cm}
    \label{fig:ods}
\end{figure}

The Cross layer that we use is a slight modification of the basic Cross layer from Section 3.2, \cite{Wang2021}. Although the actual implementation of the DCN v2 Cross layer, see Section 3.5 in \cite{Wang2021}, offers more efficient neural networks via it's low-rank mixture architecture, it exhibits information loss due to the projective nature of subsequent applications of the layer -- each Cross application first projects the input to a lower dimension, where it performs the cross operation\footnote{Formal argument for proving exact information loss bounds via Rank-Nullity Theorem is beyond the scope of this paper.}. We noticed that we can achieve similar, or better predictive performance with fewer Cross layers, which perform explicit feature crosses. We refer to this iteration of the Cross layer as the \emph{onlydense} layer\footnote{Name stems from the observation that dense layers with input coupling yield sufficient performance.}, as it intuitively performs all crosses in the intrinsic dense space (not projected). The low-rank Cross layer's formulation is as follows~\cite{Wang2021},
\begin{align*}
    x_p &= \alpha(\textbf{W}_0 \cdot \textbf{x} + b_0) \\
    x_o &= \textbf{W}_1 \cdot x_p + b_1  + \phi \cdot \textbf{x} \\
    x_r &= x_0 \odot x_o + \textbf{x},
\end{align*} where $W_0 \in \mathbb{R}^{d\times p}, W_1 \in \mathbb{R}^{p\times d}$ and $d$ and $p$ denote the output and the projection dimensions, respectively. In contrast, the \emph{onlydense} counter-part presented in this work is defined as
\begin{align*}
x_t &= \alpha(\textbf{W} \cdot \textbf{x} + b_0 ) \\
x_r &= x_t \odot \textbf{x} \cdot \phi.
\end{align*}
Here, $\alpha$ denotes the activation function (ReLU in practice) and $\textbf{W} \in \mathbb{R}^{d\times d}$. The equation represents an \emph{explicit} cross projection operation, followed by an anchor mapping achieved via a scaled Hadamard product\footnote{In practice $\phi$ has values between 1.0 and 3.0.}.
Note that the proposed layer does not perform element-wise summation, but performs only multiplication of the inputs with activated weight space of the same dimension, further scaled via a factor. Note also the absence of $x_o$ - the anchor multiplication considered by DCNv2 Cross layer -- DCN$^2$ operates not by anchoring each Cross operation with initial embedding values $x_0$, but rather iterative multiplication in the same dimension\footnote{Note that for the case of a single layer, we can assume $x = x_0$.} (Figure~\ref{fig:ods}).
In terms of computational complexity, the \emph{onlydense} layer is $O(d^2)$, same as the basic Cross layer from \cite{Wang2021}, and thus obviously heavier than the low-rank Cross layer, which is $O(d \cdot p)$. However, we observed a trade-off between computational complexity and the expressiveness of \emph{onlydense}: The latter, as shown in offline and online experiments in the remainder of this paper, offers practical benefits, making DCN$^2$ a suitable candidate for large-scale production\footnote{The inequality that governs the trade-off is $  l_{od} \cdot (|F| \cdot d) \leq l_c \cdot ((|F| \cdot d) / d_{\textrm{Cross}})$; $F$ denotes feature set $l_{\textrm{od}}$ \emph{onlydense} layer count, and $d$ dimension (left-hand side is the \emph{onlydense} layer.}.

\begin{table*}[htb!]
\caption{Performance of DCN$^2$ on benchmark data sets (AUC aggregates, window of 20k). Highlighted rows show top-performing approach w.r.t. average AUC. Highlighted are top-performing algorithms (or the ones with lowest values in case of min/std).}
\resizebox{\textwidth}{!}{%
    \centering
    \setlength{\tabcolsep}{4.5pt} 
\renewcommand{\arraystretch}{0.3} 
     \footnotesize
    \begin{tabular}{llrrrrr|llrrrrr}
\toprule
\multicolumn{6}{c}{Avazu\vspace{0.1cm}} & & \multicolumn{6}{c}{Criteo} \\
Algorithm & avg & median & max & min & std & & Algorithm & avg & median & max & min & std \\
\midrule
FM & 0.7748 & 0.7746 & 0.8304 & 0.7237 & \cellcolor{gray!20} 0.0180 & & FM & 0.7834 & 0.7831 & 0.8166 & \cellcolor{gray!20} 0.7617 & 0.0064 \\
deepFM & 0.7812 & 0.7814 & 0.8350 & \cellcolor{gray!20} 0.7230 & 0.0183 & & deepFM & 0.7906 & 0.7904 & 0.8214 & 0.7716 & \cellcolor{gray!20} 0.0063 \\
DCNv2 & 0.7826 & \cellcolor{gray!20} 0.7832 & 0.8351 & 0.7244 & 0.0183 & & DCNv2 & 0.7922 & 0.7918 & 0.8229 & 0.7730 & 0.0063 \\
\textbf{DCN$^2$} & \cellcolor{gray!20}0.7846 & 0.7846 & \cellcolor{gray!20} 0.8387 & 0.7284 & 0.0183 & &  \textbf{DCN$^2$} & \cellcolor{gray!20}0.7933 & \cellcolor{gray!20} 0.7930 & 0.8231 & 0.7751 & 0.0063 \\
\textbf{DCN$^2$}-simk & 0.7824 & 0.7826 & 0.8354 & 0.7242 & 0.0183 & & \textbf{DCN$^2$}-simk & 0.7922 & 0.7919 & \cellcolor{gray!20} 0.8233 & 0.7738 & 0.0063 \\
\midrule
\multicolumn{6}{c}{KDD2012\vspace{0.1cm}} & & \multicolumn{6}{c}{iPinYou} \\
Algorithm & avg & median & max & min & std & & Algorithm & avg & median & max & min & std \\
\midrule
FM & 0.7547 & 0.7545 & 0.8336 & \cellcolor{gray!20} 0.6769 & \cellcolor{gray!20} 0.0201 & & FM & 0.7521 & 0.7572 & 0.9955 & 0.3638 & 0.1049 \\
deepFM & 0.7719 & 0.7677 & 0.8709 & 0.7058 & 0.0260 & & deepFM & \cellcolor{gray!20}0.7669 & \cellcolor{gray!20} 0.7683 & 0.9961 & 0.4275 & \cellcolor{gray!20} 0.0997 \\
DCNv2 & 0.7730 & 0.7684 & 0.8731 & 0.7133 & 0.0265 & & DCNv2 & 0.7659 & 0.7667 & 0.9975 & 0.4333 & 0.1001 \\
\textbf{DCN$^2$} & \cellcolor{gray!20}0.7747 & \cellcolor{gray!20} 0.7699 & 0.8735 & 0.7051 & 0.0272 & & \textbf{DCN$^2$} & 0.7561 & 0.7615 & \cellcolor{gray!20} 0.9984 & \cellcolor{gray!20} 0.3574 & 0.1023 \\
\textbf{DCN$^2$}-simk & 0.7733 & 0.7693 & \cellcolor{gray!20} 0.8761 & 0.7105 & 0.0266 & & \textbf{DCN$^2$}-simk & 0.7467 & 0.7518 & 0.9980 & 0.4181 & 0.1043 \\
\bottomrule
\end{tabular}}
\label{sec:perf}
\end{table*}
\subsection{Making pairwise interactions explicit}
Field-aware Factorization Machines (FFMSs) are amongst the most expressive types of architectures (especially their deep counter-parts). However, integrating them directly alongside Deep/Cross layers incurs a substantial computational overhead, and is in practice infeasible. To remedy this shortcoming, we propose a substantially simplifed version of the FFM idea, inspired by ~\cite{10.1145/3178876.3186040}, vectorized in the form of a single dot product + projection at the pairwise interaction level (SimLayer in Figure 1). We observed that including this layer complements the Deep/Cross part, indicating that the algorithmic bias of this layer is complementary to that of DCNv2. The layer can be formulated as
\begin{align*}
    \hat{y}_{\textrm{sk}} = \alpha\left( \sum_{i=1}^{n} \sum_{j=1}^{n} \textbf{w}_{k'(i,j)} \left( \sum_{k=1}^{m} \textbf{e}_{ik} \cdot \textbf{e}_{jk} \right) + b \right),
\end{align*}
where $\textbf{e}_{ij}$ corresponds to an indexed input embedding, and $\textbf{w}_{k'(i,j)}$ to the mapping from matrix indices to the flattened vector index ($m$ and $n$ denote representation dimensions). The output activation is denoted by $\alpha$.
The layer directly models pairwise interactions between embeddings via (activated) dot-product based similarity calculation. Albeit \emph{redundant} (there are two representations per interaction), we noticed directly multiplying the matrices is faster than doing slicing/sub-setting on the fly. Over-parametrizing each interaction also showed no loss in predictive performance w.r.t. single-parameter implementation. The layer is conceptually aligned with the FFM idea in the sense that it explicitly computes feature pairs via factorization, even though it diverges when considering the activation part. Finally, we noticed that the best way to integrate this layer with the existing DCN stack is to include it as an additional \emph{logit} -- directly into the final layer of the neural network as
    $\hat y_f  = \sigma(\hat y_{dcn} + \hat y_{sk} + b_f)$,
\noindent here, $\hat y_f$ denotes the final prediction, $\hat y_{dcn}$ the output of the DCN part of the architecture and $\hat y_{sk}$ the output of the similarity layer. The $b_f$ represents the bias term and $\sigma$ the sigmoid activation.
\begin{figure}[b!]
    \centering
      \vspace{-0.4cm}
    \includegraphics[width=1.0\linewidth]{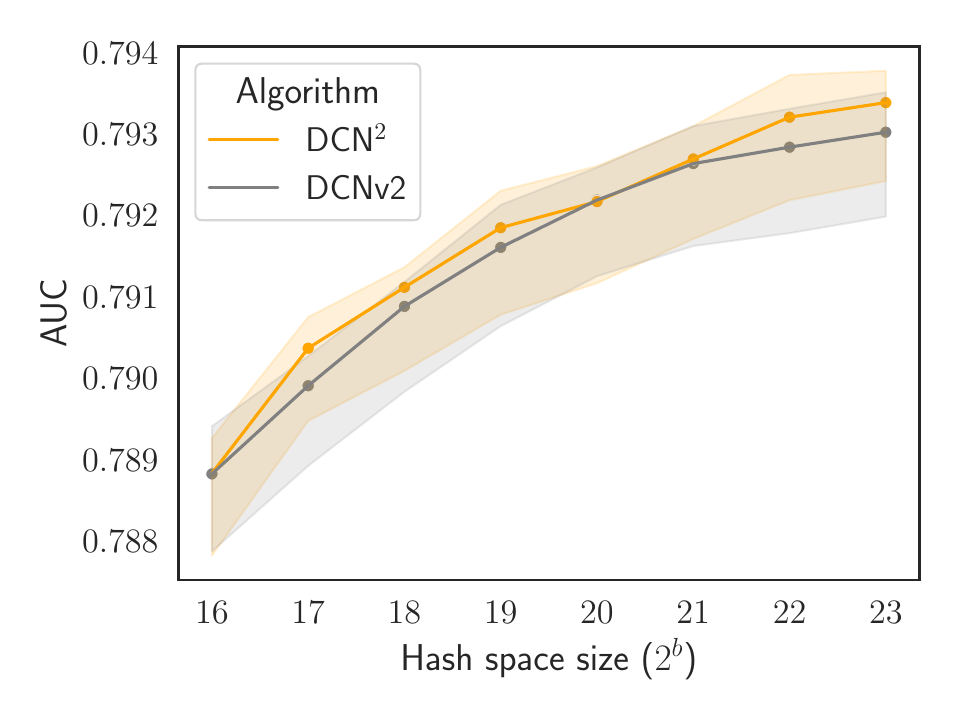}
      \vspace{-1cm}
    \caption{Expected AUC behavior (20k windows of instances) with regards to size of the hash space (Criteo).}
    \label{fig:hashs}
\end{figure}
\section{Benchmarks on open data sets}

We proceed by presenting the performance on known data sets. The experimental setting was conducted as follows.
We considered the following data sets; Criteo\footnote{\url{https://www.kaggle.com/c/criteo-display-ad-challenge}.}, Avazu\footnote{\url{https://www.kaggle.com/c/avazu-ctr-prediction/data}.}, iPinYou~\cite{zhang2014real} and KDD2012\footnote{\url{https://www.kaggle.com/c/kddcup2012-track2}}). Log transform of continuous features was conducted with no additional data pruning (rare values etc.) (as is done in our system)\footnote{Such minimal pre-processing is within reach of a regular production.}.
For each of the four benchmark data sets we considered minimal preprocessing as done in related work~\cite{vskrlj2024bag}. For each data set, we consider only \emph{single pass} learning, as this type of training is the most reflective of actual production performance discussed in the following sections. Batch size considered was 2500, for each algorithm-data set pair, we varied learning rate (range from 0.0001 to 0.01), Adam's "beta one" parameter (range 0.0 to 0.9) and embedding dimension (from 8 to 16). For increased readability, we report best-performing configurations for data set-algorithm pairs. Evaluation is conducted by computing metrics of interest in time windows of 20,000 instances. Each experiment was repeated 3 times to ensure consistency (seed level, single pass)\footnote{Averages are also done at window level to obtain the reported results.}. The hashing part is intentionally done with an internal hashing engine (same is used in production), where murmur3 hash function is considered.

Overview of AUC-based performance is shown in Table~\ref{sec:perf}. It can be observed that on average, DCN$^2$ is competitive to DCNv2 counterpart. The result indicates that DCN$^2$ can achieve superior performance to DCNv2 with approximately similar amount of hyperparameter optimization (budget for AutoML search was fixed to one hour per algorithm).
\begin{figure*}[htb!]
    \centering
    \begin{subfigure}{.25\linewidth}
      \centering
      \includegraphics[width=\linewidth]{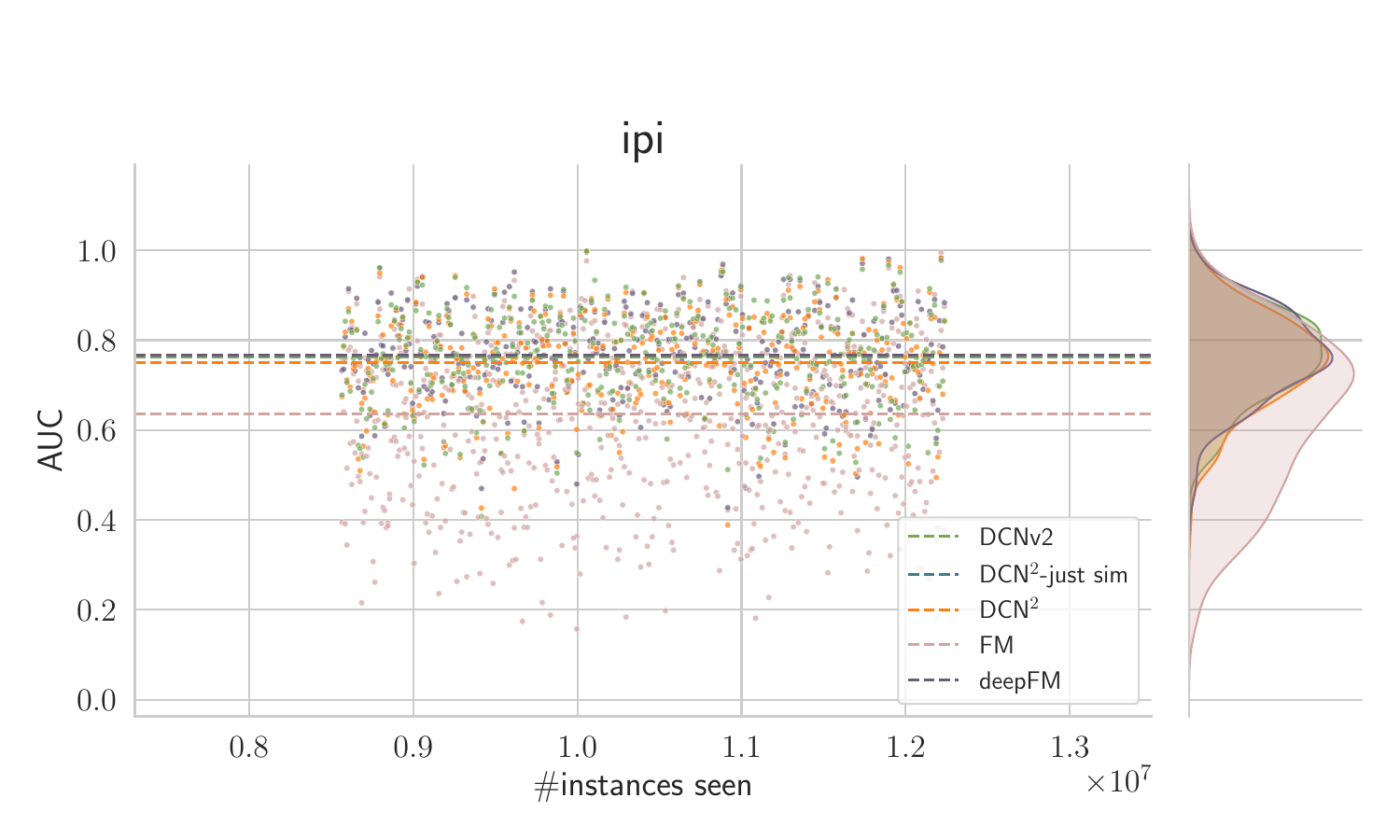}
      \caption{iPinYou}
      \label{fig:sub1}
    \end{subfigure}%
    \hfill
    \begin{subfigure}{.25\linewidth}
      \centering
      \includegraphics[width=\linewidth]{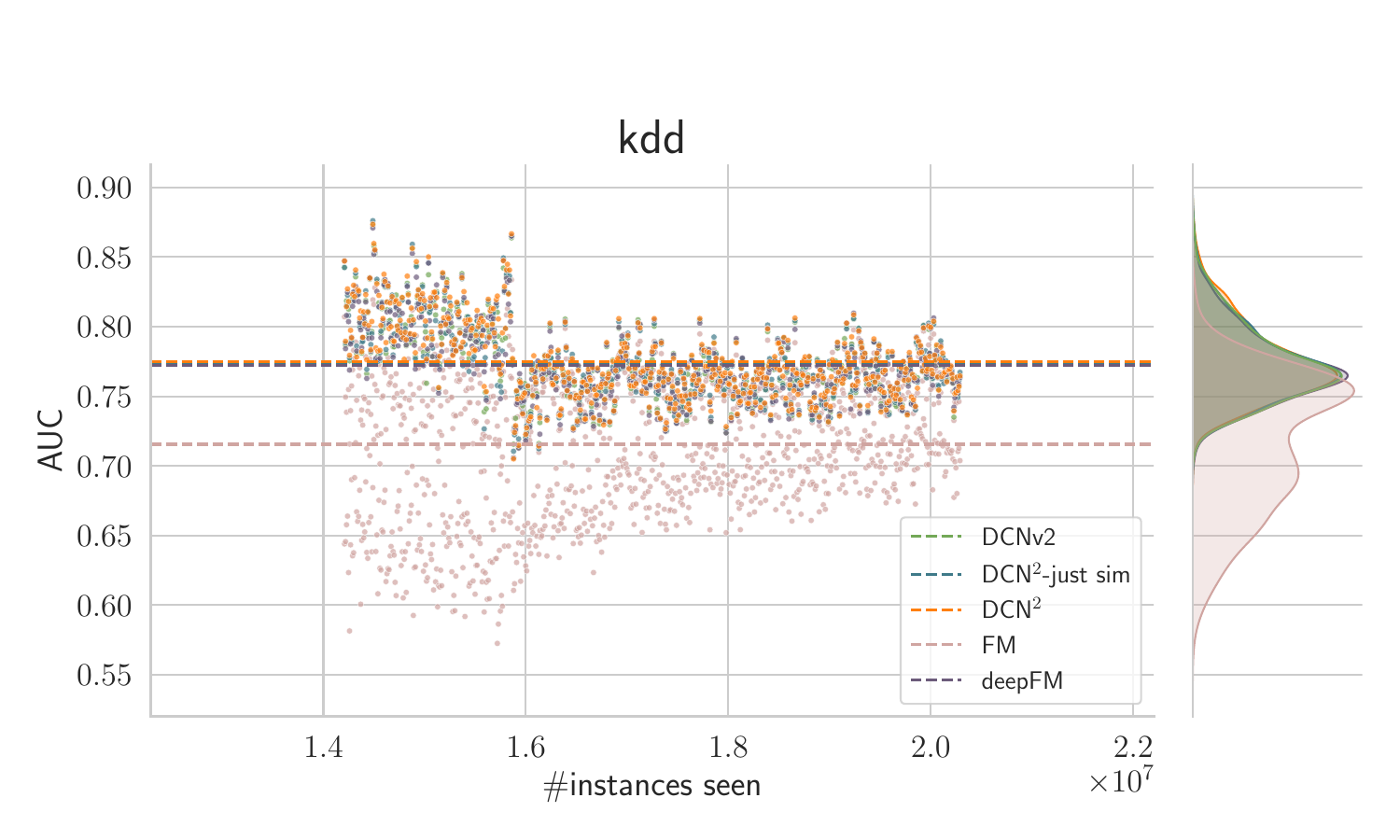}
      \caption{KDD2012}
      \label{fig:sub2}
    \end{subfigure}%
    \hfill
    \begin{subfigure}{.25\linewidth}
      \centering
      \includegraphics[width=\linewidth]{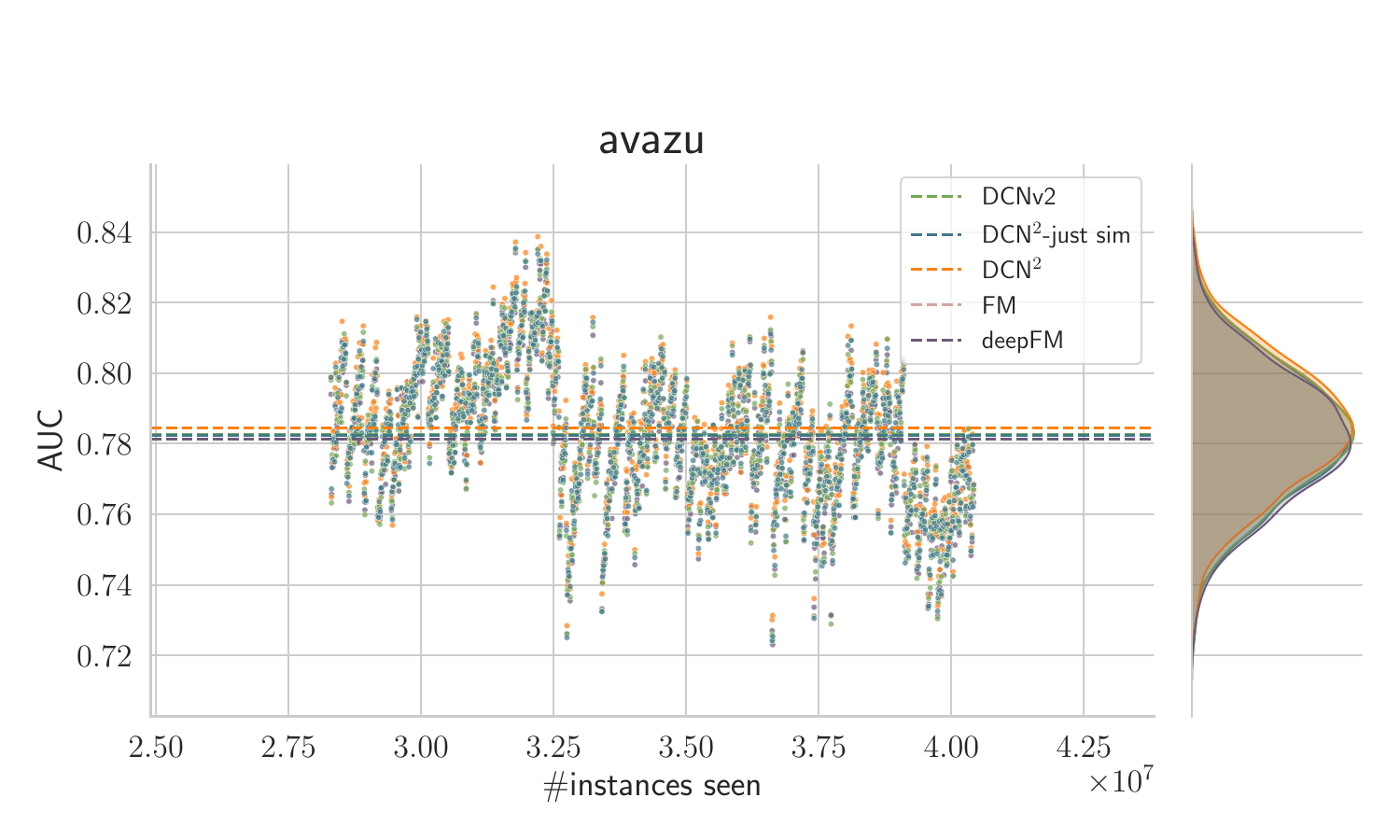}
      \caption{Avazu}
      \label{fig:sub3}
    \end{subfigure}%
    \hfill
    \begin{subfigure}{.25\linewidth}
      \centering
      \includegraphics[width=\linewidth]{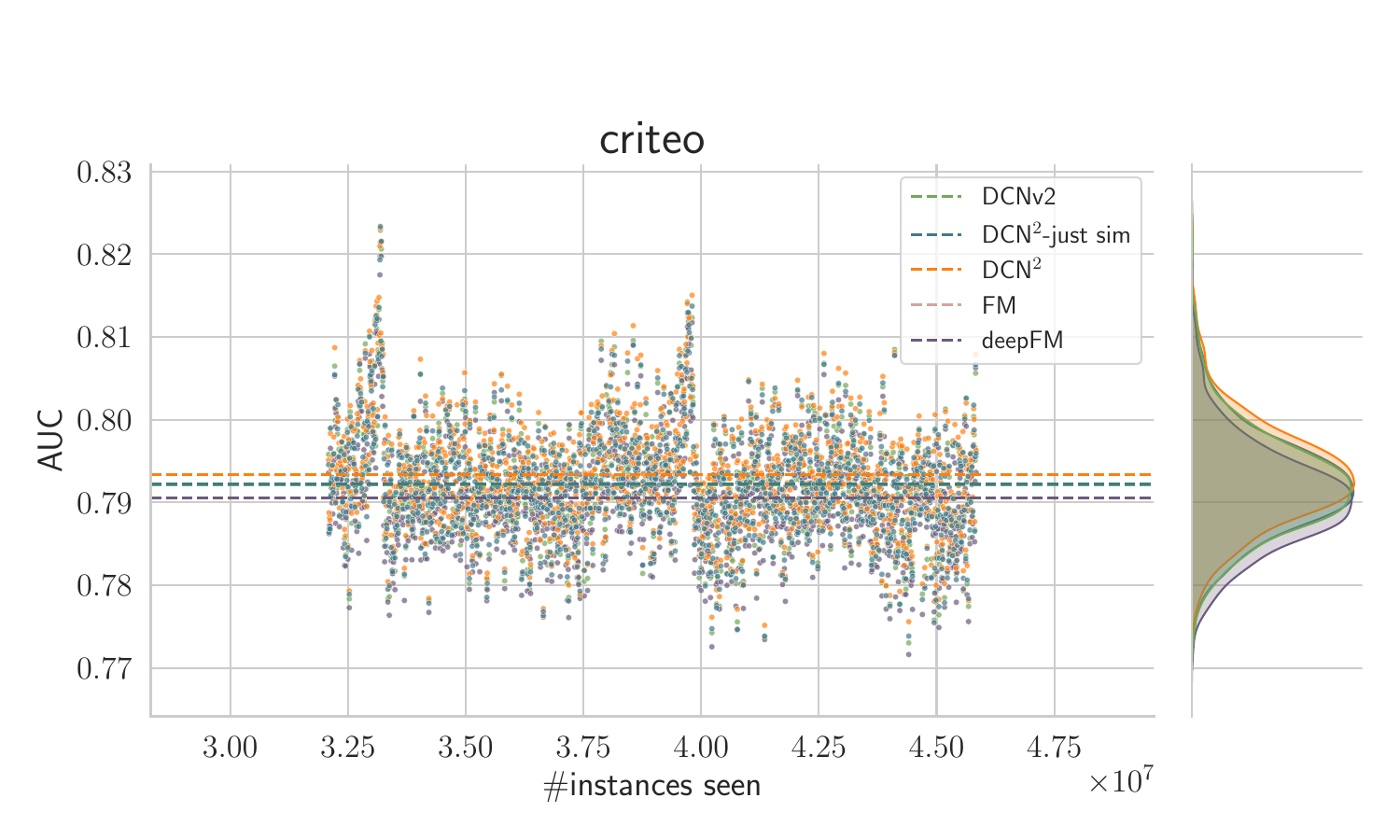}
      \caption{Criteo}
      \label{fig:sub4}
    \end{subfigure}%
    \vspace{-0.1cm}
    \caption{Visualization of overall performance of different algorithms (single-pass) across different benchmark data sets. Kernel estimates at the rightmost part of each visualization denote expected performance across the whole data stream snapshot.}
    \label{fig:auc-benchmark}
\end{figure*}
We next present incremental learning traces of main algorithms. Trace-level analysis is useful to better understand convergence properties and overall variability of performance metrics during the whole learning lifetime. Visualization of the performance traces for all data sets and algorithms of interest, is shown in Figure~\ref{fig:auc-benchmark}. It can be observed that the proposed DCN$^2$ consistently out-performs the baseline approach (also confirmed by the drift of the fitted gaussian at the rightmost part of the image). Results indicate that either DCN$^2$ or the similarity kernel-only (\emph{-simk}) version of the architecture offer competitive performance. It can also be observed that performance of approaches is different for the iPinYou data set, which most likely points at different data regime for this problem.

\section{Hash space scaling laws}
To understand the hypothesized (and measured) positive impact of collision-weighted lookups, we conducted additional ablation study, where we computed models for a range of hash spaces of different sizes, with the goal of identifying possible consistent properties of the performance (e.g., RIG) metric w.r.t. size of the space. For this experiment, we varied hash space size of the best (on average) performing configurations of the models. Overview of the experiment is shown in Figure~\ref{fig:hashs}.
It can be observed that the DCN$^2$ architecture almost consistently out-performs the DCNv2 baseline. Further, the relation between the RIG and hash space size is of quasi-linear nature, indicating that with smaller hash space, performance drops for all approaches, even though less for collision-weighted lookups; this result denotes architecture's behavior with regards to different collision rates (the only thing that changes with varying hash space size). Even though collision-weighted lookups are presented as intrinsic part of DCN$^2$, they are agnostic to down-stream architecture, and can be used with any lookup-based approach.

\begin{figure}[b!]
    \centering
    \vspace{-0.4cm}
    \includegraphics[width=1.0\linewidth]{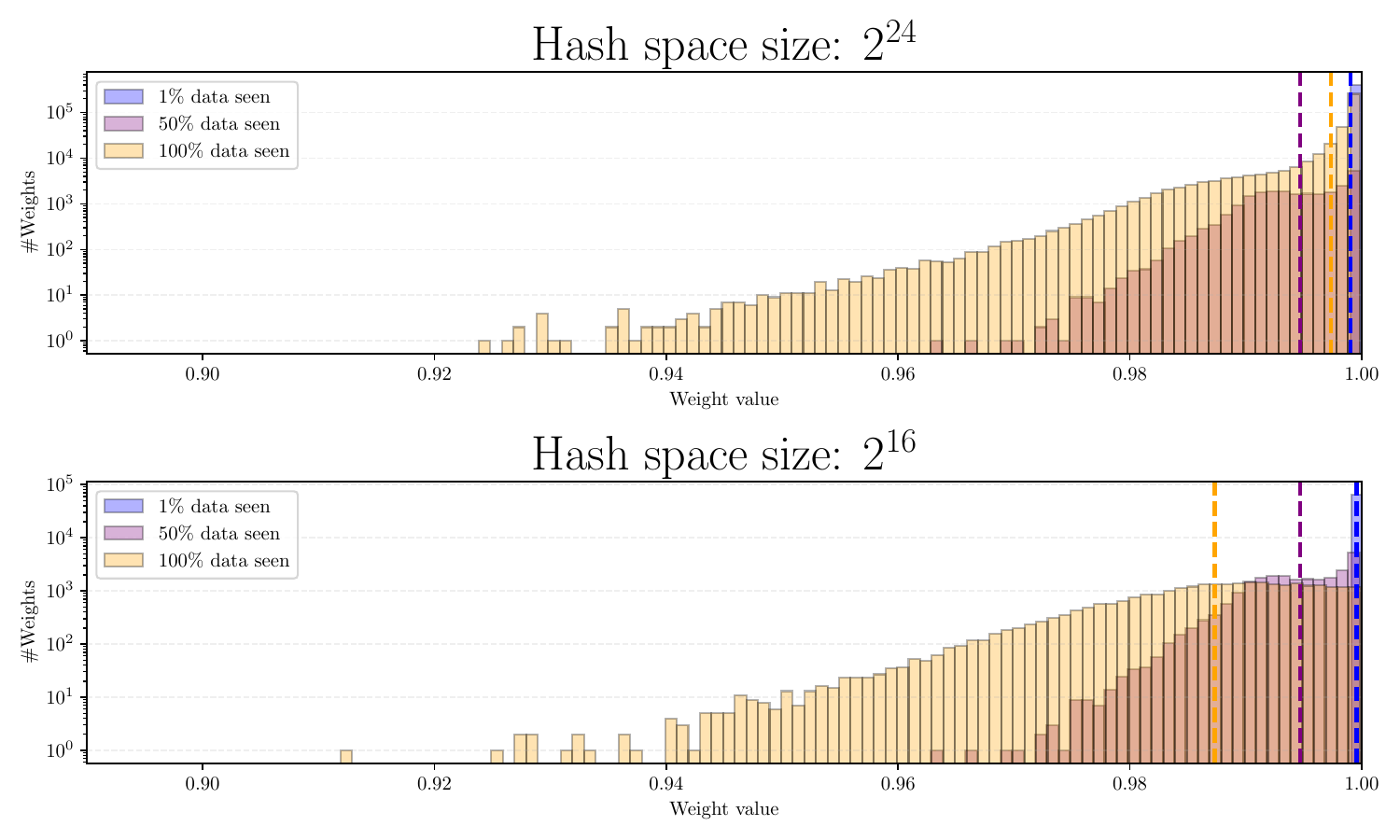}
    \vspace{-0.9cm}
    \caption{Distribution of collision weights after a single pass on the KDD2012 data set.}
    \label{fig:col-weights}
\end{figure}

We proceed by discussing the \textbf{distribution of collision weight space}.
In this paper we posit that collision weights change during training (from initial value of 1.0). It's likely there are not many collision-points that would require considerable modification, yet to further study this phenomenon, we computed distribution of collision weights after training a DCN$^2$ model on the KDD data set. It can be observed that most of the modified weights are down-scaled (value smaller than one), indicating the network learns to silence only selected lookups. There is a distinct heavy-tailed distribution governing this effect -- only a few lookups are substantially down-scaled, most are only partially modified (if at all). The distribution of the resulting weights is shown in Figure~\ref{fig:col-weights}. With smaller hash space, more smaller weights are observed (averages denoted with dotted lines).
\section{DCN$^2$ online: 0.5 billion+ predictions per second and A/B results}
The presented DCN$^2$ is used for multiple use cases, ranging from conversion rate prediction to click-through rate prediction. It has been scaled to a point it produces more than 0.5 billon predictions per second, and has been extensively A/B tested. This section focuses on online performance - i.e. behavior of the whole system that depends on DCN$^2$ for inference. KPI-wise, we considered offline and online lifts -- offline metrics of interest are relative information gain (RIG), and online are Revenue per Session and mille (RPS/RPM), gross revenue, CPA and conversion ratio (see footnote 11 for details). On both ends, the DCN$^2$ has substantially outperformed existing DCNv2-based system/implementation. End-to-end evaluation is shown next\footnote{Result shows aggregate lift of many A/B tests conducted as part of rolling out the architecture. swCR=spend-weighted conversion ratio and GR=Gross revenue.  swCR is the ratio of the number of conversions in the variant to the number in the control. A ratio > 1 indicates improved conversion volume. For CVR test, we also observed -2.8\% CPA, the ratio of the Cost per Action (CPA) under the variant to the CPA under the control.  A CPA ratio < 1 indicates improved cost efficiency..}.
\begin{table}[htb!]
    \centering
   \vspace{-0.3cm}
    \begin{tabular}{c|c|c}
         Use case & Lift Offline (AutoML)  & Lift Online (A/B) \\ \hline
         CTR & 0.0035 (RIG) & 3.2\% RPM \\
         CVR & 0.0010 (RIG) & 4.2\% swCR, 0.37\% GR
    \end{tabular}
      \vspace{-0.3cm}
    \label{tab:online}
\end{table}

To achieve the mentioned scale, we optimized multi-value feature integration through three enhancements: Summation-Based Embedding Aggregation condenses variable-length multivalue inputs into fixed-size embeddings using summation, balancing expressiveness and computational efficiency. Fixed-Length Tensors with Padding not only introduce uniformity for batch processing and vectorized operations by padding feature values to set lengths but also boost throughput, though careful padding length choice is necessary to maintain model fidelity. Data-Driven Padding Optimization automates padding length decisions via a streaming statistics estimator, using historical training data to derive padding length from empirical feature cardinality distributions, ensuring adaptability to evolving data without manual intervention.

\section{Migrating from DCNv2 - an AutoML case study for win probability models}
In this section, we present a possible approach to migrating an existing model from DCNv2 to DCN$^2$ by discussing our AutoML case study for win probability models. Here, we refer to the AutoML as the automated process for architecture search, hyper-parameter optimization (HPO) and feature selection w.r.t. the target performance metric (RIG) in an offline mode. Beyond the change in architecture, there are two other significant novelties in the migrated model. That is, the embedding dimension increased from 6 in DCNv2 to 16 in DCN$^2$, and, unlike the initial model, the proposed model does not include any feature combinations --- we avoided those on purpose to demonstrate the effectiveness of the onlydense layer. Introducing these improvements resulted in significant offline lift of 0.01552 (RIG).

\section{DCN$^2$ Inference - lessons and battle scars}

Our production pipeline was progressively re-engineered to deliver >0.5 billion DCN² predictions per second while sustaining strict latency budgets.
Model inference first ran in a proprietary FwumiousWabbit service~\cite{vskrlj2024bag}. To benefit from an industry-standard ecosystem we ported the model to TensorFlow, then temporarily to ONNX, whose runtime initially offered lower p99 latency at equal CPU cost. Subsequent kernel-level and graph optimizations allowed the TensorFlow path to match ONNX, so the final deployment now supports either backend.
Custom TensorFlow and ONNX builds linked against OpenBLAS and Intel MKL were benchmarked on production hardware; none outperformed the stock binaries, so default builds were retained.
A “local fan-out’’ algorithm replaces conventional dynamic batching. Each large request is partitioned into fixed-size micro-batches that share pre-allocated input tensors, eliminating per-request buffer creation. The scheme minimizes allocator traffic while keeping CPU/prediction within 3 \% of the theoretical optimum.

\begin{figure}[t!]
    \centering
     \fbox{\includegraphics[width=\linewidth]{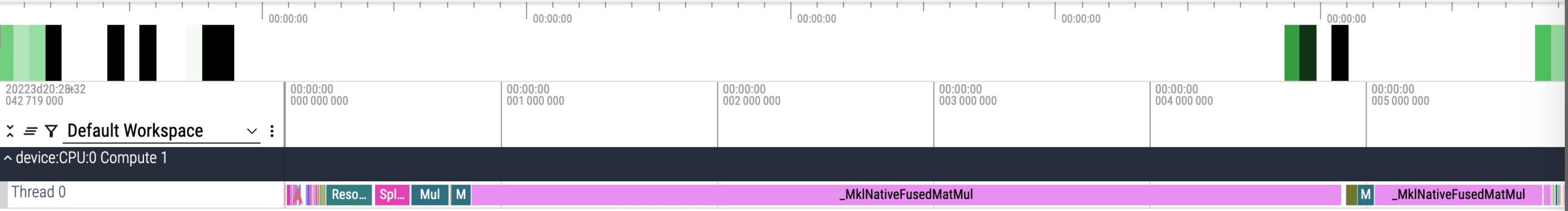}}
    \caption{Inspecting introduced layers (pink) during inference.}
    \label{fig:fused}
    \vspace{-0.5cm}
\end{figure}

TensorFlow intra-op and inter-op threads were pinned to the actively used cores, and the runtime’s background spin-wait was disabled. Removing this surplus parallelism cut scheduler contention and reduced p99 latency by 18 \%.
All nonessential computations were moved off the critical path and executed asynchronously via lightweight queues. Memory traffic was further reduced by reusing feature and output buffers across calls; Jemalloc replaced the default allocator to curtail fragmentation and GC overhead. Combined, these service-layer changes improved throughput by 1.6× while holding tail-latency constant.
Figure \ref{fig:fused} illustrates the final fused operator profile obtained from online instrumentation.
Through a relentless focus on optimizations across multiple layers, we successfully achieved the scale, stability, and performance required for a production-grade system. These iterative enhancements enabled us to meet the challenge of processing more than 0.5 billion predictions per second, all while upholding strict latency requirements and maintaining peak operational efficiency.

\section{Conclusions}

DCN$^2$ consistently outperforms baseline recommenders both offline (e.g., higher AUC on Criteo and Avazu) and online (RPM gains in production). Its collision-weighted look-ups mitigate feature collisions in high-dimensional sparse data, yielding richer embeddings and better predictions. The lightweight \emph{onlydense} layer replaces costly explicit crosses with an explicit interaction term, improving accuracy at lower compute cost. In production DCN² drives measurable revenue lift and sustains >5×10$^8$ pps while running on standard ML stacks. Further, the models produced from scratch by using $DCN^2$ contain minimal number of explicit (hashed) interactions, substantially simplifying the final production-ready models. Future work includes investigation of more aggressive collision weight schemes and LoRA based fine-tuning of existing models (with implications for transfer learning).
\bibliographystyle{acm}
\bibliography{bibfile}

\end{document}